\documentclass[sigplan,twocolumn]{acmart}
\renewcommand\footnotetextcopyrightpermission[1]{}
\acmConference[Arxiv]{}{2025}{}

\usepackage{tcolorbox}
\begin{document}
\title{H2: Towards Efficient Large-Scale LLM Training on Hyper-Heterogeneous Cluster over 1,000 Chips}

\author{Ding Tang}
\affiliation{%
  \institution{Shanghai Artificial Intelligence Laboratory}
  \city{Shanghai}
  \country{China}
}
\email{tangding@pjlab.org.cn}

\author{Jiecheng Zhou}
\affiliation{%
  \institution{Shanghai Artificial Intelligence Laboratory}
  \city{Shanghai}
  \country{China}
}
\email{zhoujiecheng@pjlab.org.cn}

\author{Jiakai Hu}
\affiliation{%
  \institution{Shanghai Artificial Intelligence Laboratory}
  \city{Shanghai}
  \country{China}
}
\email{hujiakai@pjlab.org.cn}

\author{Shengwei Li}
\affiliation{%
  \institution{Shanghai Artificial Intelligence Laboratory}
  \city{Shanghai}
  \country{China}
}
\email{lucasleesw9@gmail.com}

\author{Huihuang Zheng}
\affiliation{%
  \institution{Shanghai Artificial Intelligence Laboratory}
  \city{Shanghai}
  \country{China}
}
\email{zhenghuihuang@pjlab.org.cn}

\author{Zhilin Pei}
\affiliation{%
  \institution{Shanghai Artificial Intelligence Laboratory}
  \city{Shanghai}
  \country{China}
}
\email{peizhilin@pjlab.org.cn}

\author{Hui Wang}
\affiliation{%
  \institution{Shanghai Artificial Intelligence Laboratory}
  \city{Shanghai}
  \country{China}
}
\email{wanghui@pjlab.org.cn}

\author{Xingcheng Zhang}
\affiliation{%
  \institution{Shanghai Artificial Intelligence Laboratory}
  \city{Shanghai}
  \country{China}
}
\email{zhangxingcheng@pjlab.org.cn}

\begin{abstract}
Recent advancements in large language models (LLMs) necessitate extensive computational resources, prompting the use of diverse hardware accelerators from multiple vendors. However, traditional distributed training frameworks struggle to efficiently utilize hyper-heterogeneous clusters comprising thousands of chips due to significant disparities in software stacks, operator implementations, communication libraries, and hardware capabilities. To address these challenges, we propose H2, which stands for HyperHetero and is a systematic framework enabling efficient training of LLMs on clusters with over 1,000 heterogeneous chips. H2 incorporates DiTorch\footnote{https://github.com/DeepLink-org/ditorch}, a unified PyTorch-compatible interface ensuring program consistency across chips, and DiComm, a device-direct RDMA communication library optimized for heterogeneous environments. Furthermore, we introduce HeteroPP with HeteroAuto, an adaptive pipeline parallelism strategy that dynamically balances computational load, memory limitations, and communication overhead. Evaluations on a 100-billion-parameter LLM demonstrate that our approach consistently achieves a superlinear speedup, outperforming baseline homogeneous training solutions by up to 16.37\% in our experiments. These findings validate the feasibility and efficiency of hyper-heterogeneous training at unprecedented scales.
\end{abstract}

\settopmatter{printfolios=true}
\maketitle

\section{Introduction}\label{sec:Intro}

\begin{figure}
    \centering
    \includegraphics[width=0.9\linewidth]{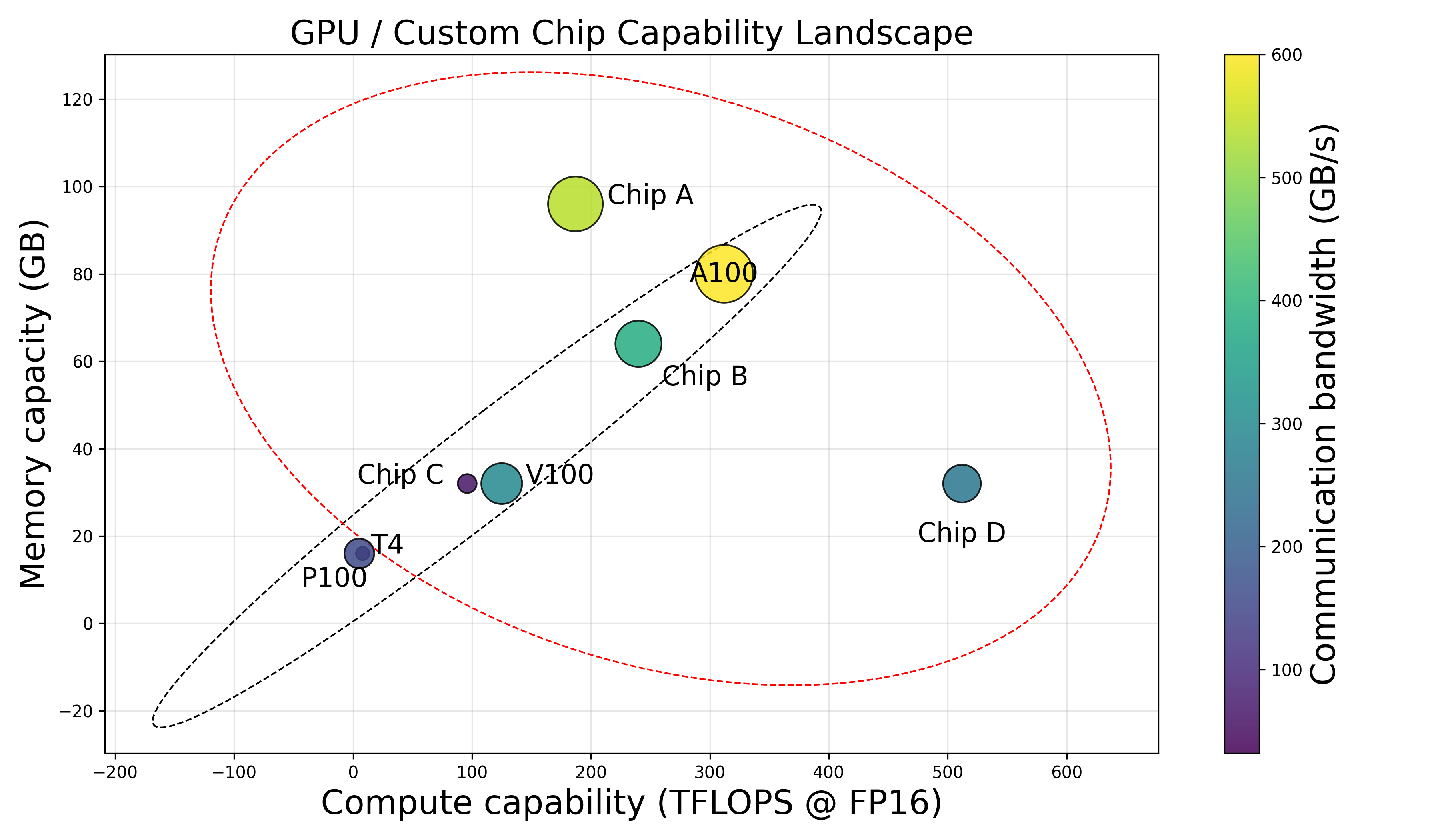}
    \caption{Comparison of chip specifications between capability-incremental and our hyper-heterogeneous scenario.In traditional heterogeneous scenarios, as indicated by the black dashed circles in the figure, chips show a trend of increasing capabilities in computation, communication, and memory. In contrast, in hyper-heterogeneous scenarios, as indicated by the red dashed circles in the figure, the capabilities of chips in these three aspects do not follow any specific pattern.}
    \label{fig:gpu_capacity}
\end{figure}

Recent advances in artificial intelligence have led to the fast development of large-scale models such as GPT-4\cite{achiam2023gpt}, LLaMA\cite{touvron2023llama}, and DeepSeek\cite{guo2025deepseek,liu2024deepseek}, which require substantial computational resources. The training of these models typically involves prolonged operation on expansive computing clusters to ensure convergence and optimal performance. Historically, such clusters have been built around single-chip servers\cite{10.1109/SC41406.2024.00089}. However, as hardware requirements continue to escalate, numerous semiconductor companies have introduced their own distinctive chip designs\cite{otterness2020amd}, and each chip is developed according to proprietary design principles, resulting in both \textbf{software isolation and different hardware specifications}.

The real-world application of heterogeneous chips is stead-ily becoming a reality. First, relying solely on a single type of chip may be insufficient due to limitations in production capacity or the specific suitability for certain computational tasks\cite{milojicic2021future}. Consequently, many organizations are now deploying heterogeneous clusters composed of multiple types of chips\cite{276938}. Furthermore, with the ongoing procurement of new chips over time\cite{elster2022nvidia, tirumala2024nvidia, choquette20213}, it is urgently important to integrate legacy chips with newer ones efficiently to reduce idle costs.

\textbf{Hyper-heterogeneous} computing refers to an emerging computing paradigm characterized by significant hardware and software diversity within large-scale computing clusters. Specifically, hyper-heterogeneous environments exhibit three key distinguishing features. First, the hardware specifications—including computational power, memory capacity, and communication bandwidth—of different chips can vary drastically without clear patterns or predictable relationships. Second, chips from different vendors typically exhibit pronounced software isolation, utilizing separate software stacks and distinct communication libraries, thereby complicating interoperability and software integration. Third, the quantities of different chip types within a single cluster may be highly imbalanced, further increasing complexity in workload scheduling and resource management. Unlike conventional heterogeneous computing systems, which usually involve fewer chip variants with more uniform programming models and communication protocols, hyper-heterogeneous systems are characterized by a significantly higher degree of complexity and integration challenges.

Existing heterogeneous approaches have predominantly concentrated on designing training algorithms for hardware setups that often involve different generations of chips from the same vendor\cite{jia2022whale, um2024metis}. For example, Metis searches for heterogeneous parallelism strategies on Nvidia T4, P100, and V100, and Whale accelerates heterogeneous parallel training with Nvidia P100 and V100. This design choice leads to two notable implications. First, there is an inherent performance gap between pairs of chips; one type typically outperforms the other across a range of specifications such as computational power, memory capacity, and communication bandwidth as shown in Figure~\ref{fig:gpu_capacity}. This discrepancy can significantly impact workload distribution and overall system efficiency. Second, these setups benefit from a naturally unified software stack\cite{NVIDIAcutlass2025} and communication library\cite{github_nccl} (e.g., CUDA and NCCL), which minimizes the need for adjustments that would otherwise be required to bridge isolated software environments within heterogeneous systems. 

In addition, most heterogeneous training approaches have been designed for small-scale scenarios or validated only on clusters with a limited types of chips\cite{xu2024hethub}. Consequently, their scalability and performance in larger, more diverse computing environments remain largely unproven.

In this work, we address the scenario of efficiently training extremely large models in hyper-heterogeneous computing environments. To uniformly leverage chip resources from different vendors while ensuring scalability, we highlight the necessity of developing new systems and algorithms specifically designed for hyper-heterogeneous scenarios. Such systems must effectively address the following issues:

First, the technical isolation among different types of chips are considered from three aspects: software, communication and operators. The \textbf{software isolation} comes from the fact that chips from different vendors are supported by distinct technology stacks and, in many cases, their own training frameworks. Although some chips offer compatibility via adaptations to popular interfaces\cite{wu2023pytorch, abadi2016tensorflow} (e.g., PyTorch), semantic inconsistencies between these solutions pose significant challenges for developers aiming for a unified development environment. Furthermore, the issue of \textbf{communication isolation} adds another layer of complexity. Different chips utilize diverse internal communication libraries, and the variations in server topologies—such as the number of network interface cards and respective bandwidths—make it difficult to develop a standardized communication library that can be uniformly applied\cite{li2019evaluating}. Finally, \textbf{operator isolation} arises from the differing hardware design logics across chips. As each chip implements its own acceleration strategies and operator development methodologies, there often remain discrepancies in the support and precision of common operators. Although most chip manufacturers have implemented adaptations for widely-used large-scale models, gaps persist in both operator availability and computational accuracy alignment.

Second, heterogeneous training in \textbf{large-scale} scenarios brings several challenges. Driven by the rapid advancement of large-model techniques\cite{radford2018improving, devlin2019bert}, the computational resources necessary for training a state-of-the-art base model have grown extraordinarily demanding\cite{narayanan2021efficient}. This scale of computational requirement naturally results in an extremely large space for parallel execution strategies, making the optimization process considerably complex. Therefore, it is imperative not only to develop efficient parallel algorithms, but also to ensure that the algorithmic search process itself remains computationally efficient to reduce the high cost of large-scale computing resources\cite{zheng2022alpa, li2022amp}.

To address the two aforementioned issues, we have designed the H2 framework, which includes DiTorch, DiComm and HeteroPP. DiTorch extends PyTorch’s interface to realize user-transparent support for heterogeneous computation, requiring only a one-line code modification to the original training program, and provides a suite of tools to ensure the alignment of computational precision among different chips. DiComm implements device-direct RDMA communication between heterogeneous GPUs and optimizes communication via hardware-topology-aware methods. HeteroPP is an efficient heterogeneous training framwork which employs pipeline parallelism to connect different types of chips, and introduces an efficient automatic search algorithm for parallelism strategy. In our evaluation of a 100B LLM training on a cluster comprising 1,024 chips of four distinct architectures, our H2 framework achieved a superlinear speedup in comparison to conventional homogeneous training approaches.

In summary, our contributions are as follows:
\begin{itemize}
    \item We introduce a novel training paradigm termed hyper-heterogeneous training to address the evolving landscape of heterogeneous computing.
    \item We design and develop DiTorch and DiComm to break the technical isolation among different types of chips and achieve user-transparent heterogeneous computation and communication with only a one-line modification.
    \item We propose the HeteroPP framework, which implements highly efficient heterogeneous training algorithms, alongside HeteroAuto, a robust and efficient automatic strategy-search algorithm for HeteroPP framework under hyper-heterogeneous scenarios.
    \item We validate the efficiency and training stability of H2 system through experiments on training a 100B-parameter large model.
\end{itemize}
\section{Background}\label{sec:background}
In this section, we first present the software stack for LLM pre-training. In addition, we discuss about various parallelization strategies. Finally, we briefly introduce the characteristics of hyper-heterogeneous clusters.

\subsection{LLM Pre-training Software Stack}\label{sec:software stack}
\begin{figure}
    \centering
    \includegraphics[width=0.95\linewidth,height=7.5cm]{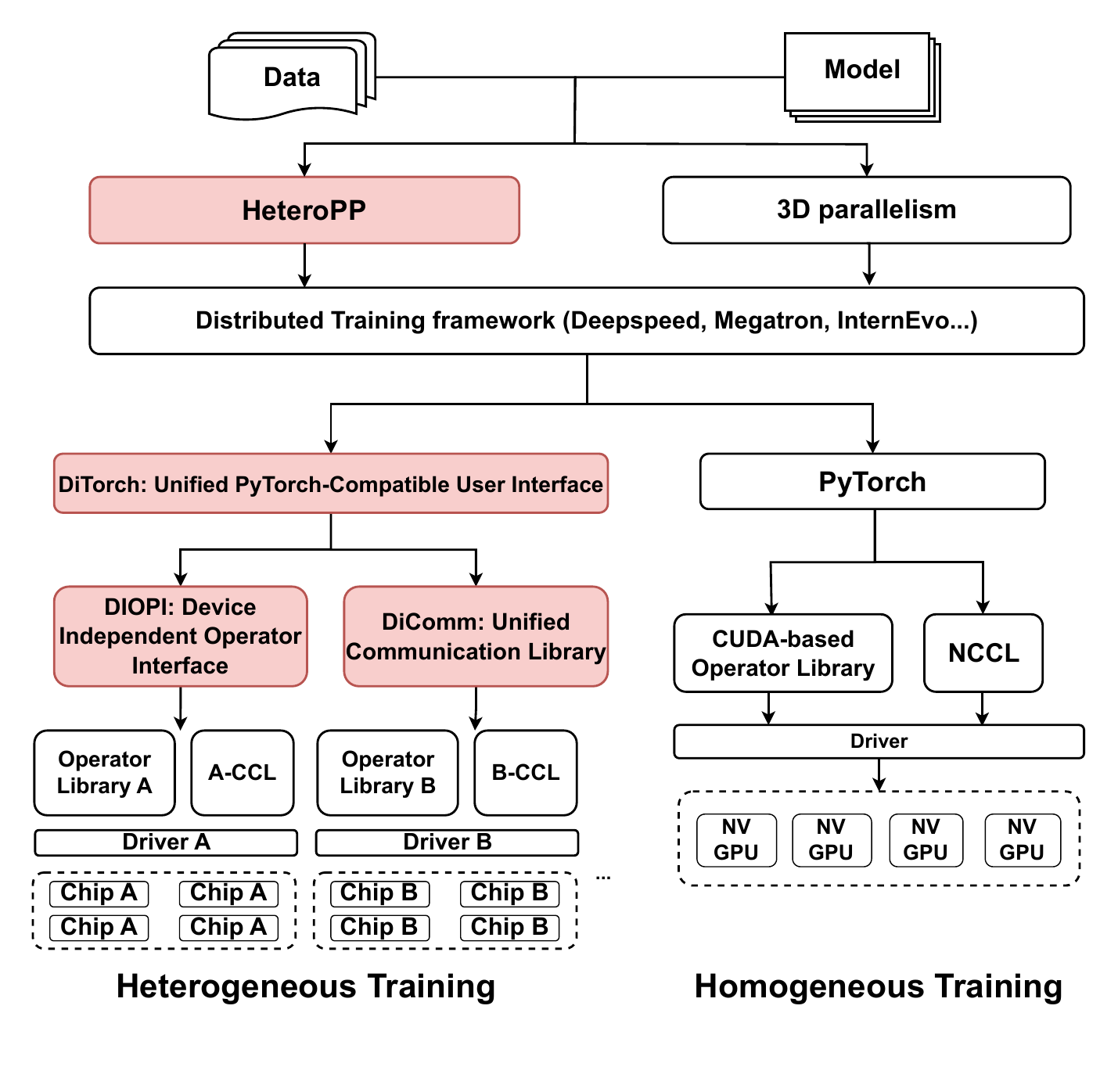}
    \caption{LLM Pre-training Software Stack. Red parts represent our work.}
    \label{fig:software_stack}
\end{figure}

Large-scale model pre-training often entails millions of iterations and involves over a hundred GPUs. As shown in Figure \ref{fig:software_stack}. first, users provide the training data and model to the pre-training framework, which efficiently implements data processing and parallelization strategies through its distributed design. Then, these distributed pre-training frameworks employ PyTorch's interfaces to realize the parallel strategies. Finally, PyTorch leverages the high-performance computing and communication libraries specific to each chip to execute the computation and communication functions.

\subsection{Distributed Training}
Our work focuses on 3D parallelism \cite{shoeybi2019megatron}, including pipeline parallelism\cite{huang2019gpipe, narayanan2019pipedream, fan2021dapple}, tensor parallelism\cite{shoeybi2019megatron, chang2024flux}, and data parallelism\cite{li2020pytorch}, which is the most commonly used parallel strategy in large model pre-training. ZeRO-1 ~\cite{rajbhandari2020zero} is enabled in default.

\textbf{Pipeline Parallelism}:
A large-scale model is split layer-wise into multiple stages, with each stage comprising several contiguous layers. These stages are then assigned to different GPUs. Pipeline parallelism works by dividing the data into multiple micro-batches, allowing the stages to operate in a pipelined fashion and thus fully leveraging the computational resources of all GPUs.

\textbf{Data Parallelism}:
In large model pre-training, training data can be distributed across multiple GPUs, with each GPU maintaining a complete copy of the model parameters to perform forward and backward computations. Once gradients are computed, an all-reduce operation is used to synchronize the parameters across GPUs.

\textbf{Tensor Parallelism}:
For large Transformer-based models, tensor parallelism is an effective method to enhance training efficiency. In this approach, parameter tensors are partitioned horizontally among multiple GPUs, thereby distributing both the memory load and the computational workload for forward and backward passes across the GPUs. When tensor parallelism is applied, each Transformer block’s forward pass requires two all-reduce operations to reconstruct the complete input values.


\subsection{Hyper-Heterogeneity Characteristics}

\begin{figure}
    \centering
    \includegraphics[width=1.0\linewidth]{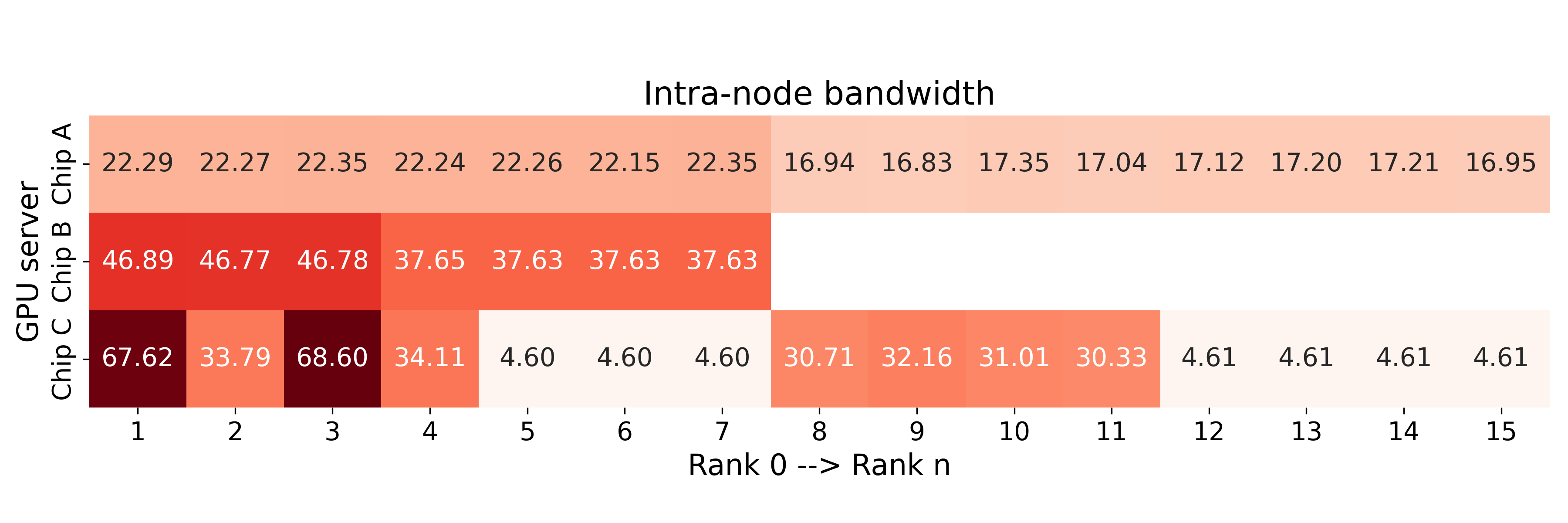}
    \caption{Intra-Node Bandwidth Performance in Different GPU Servers}
    \label{fig:intra-bdw}
\end{figure}

Training an extremely large model requires vast resources, and these resources are often needed concurrently during training. In industrial settings, computing clusters typically consist of various Chips from different vendors to meet both training and inference requirements. Utilizing heterogeneous chip resources can satisfy the demands of training tasks and experiments without having to wait for a homogeneous set of resources to become fully available, thereby accelerating model training and experimentation. Compared to homogeneous clusters and standard heterogeneous environments, Hyper-Heterogeneity has the following characteristics:

\textbf{Distinct Software Stacks for Each Vendor’s Chip}:
Due to the "isolation" between hardware design and the corresponding software stacks, the same operator implemented on different chips can produce varying degrees of precision. For example, in matrix multiplication, different vendors may employ unique data layouts and accumulation orders for operators, leading to discrepancies in the final results. Automatically detecting and resolving these precision issues is a fundamental challenge. Moreover, the software stack for large model pre-training is inherently complex, with each vendor typically focusing only on its proprietary libraries; breaking the technical “isolation” between these stacks is the cornerstone of heterogeneous parallel training.

\textbf{Heterogeneity in Computational, Communication, and Storage Capabilities Across Different Chips}:
Machines from different vendors can exhibit unpredictable differences in computing, communication, and storage capacities. For instance, as shown in Figure \ref{fig:gpu_capacity}, a Chip A might have up to 96GB of memory but only deliver 182 Tflops at FP16, whereas Chip C could achieve 512 Tflops at FP16 despite having only 32GB of memory. Such discrepancies lead to divergent choices in parallelization strategies and workload partitioning during training. Because different chips need different parallel strategies, selecting an optimal combination for large-scale, multi-vendor configurations becomes challenging. The rapid proliferation of GPU models and expanding cluster sizes call for a scalable parallel approach.

\textbf{Complex Intra-node Topologies}:
Popular GPU servers often utilize high-speed intra-node interconnections\cite{li2019inter} (e.g., NVLink) to achieve high and homogeneous bandwidth within a node. However, some GPU servers either lack or do not fully implement high-speed intra-node connections, resulting in varying communication capabilities among GPUs within the same node. This discrepancy an exponential search space, rendering the Metis algorithm extremely inefficient in hyper-heterogeneous scenarios.

\section{Breaking Software and Communication Barrier}

\begin{figure}
    \centering
    \includegraphics[width=0.95\linewidth]{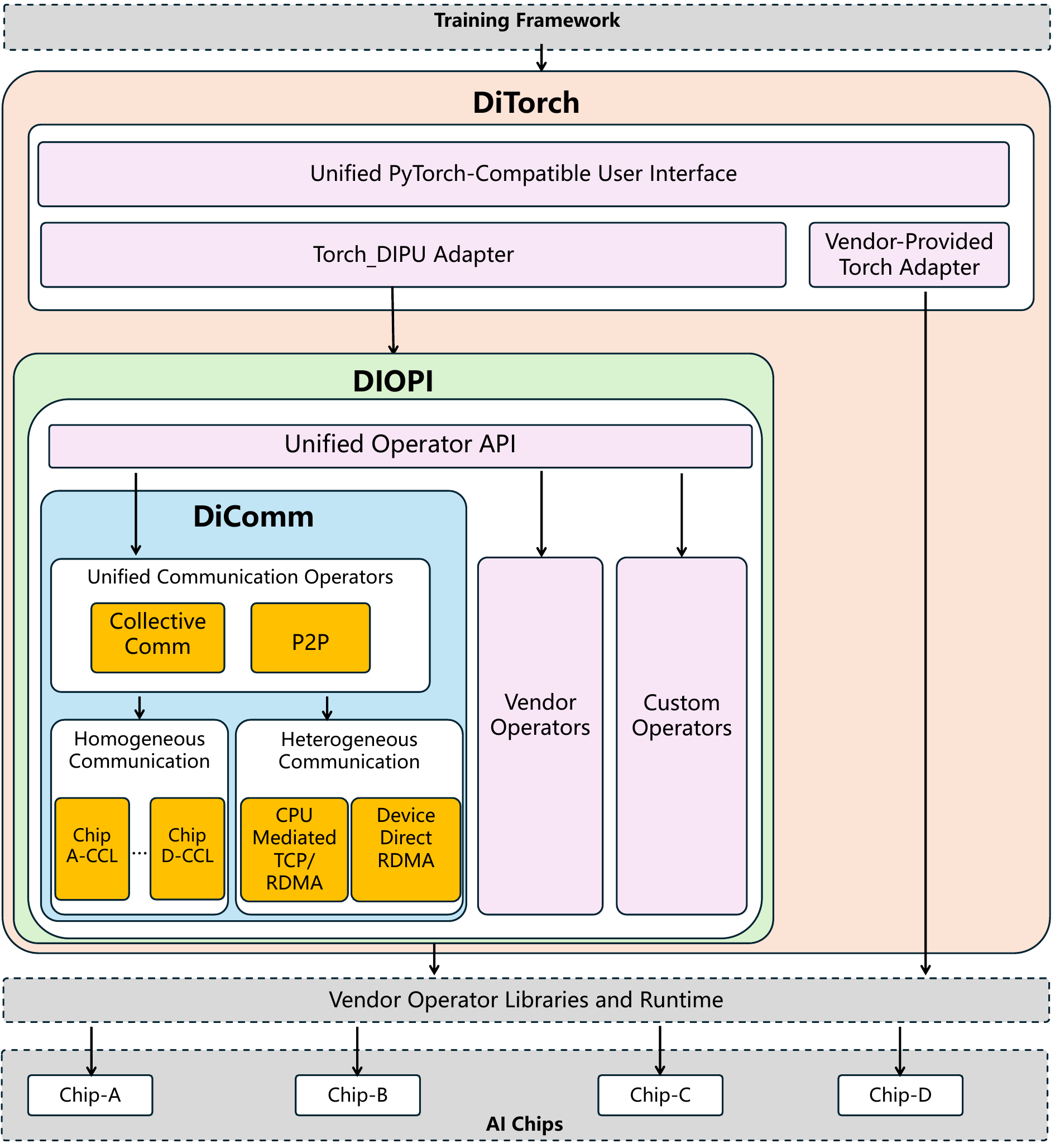}
    \caption{System Overview of DiTorch and DiComm.}
    \label{fig:ditorch_arch}
\end{figure}
To address the software and communication isolation challenges outlined previously, we introduce \textbf{\textit{DiTorch}} and \textbf{\textit{DiComm}} in this section, whose system overview are shown in Figure \ref{fig:ditorch_arch}. DiTorch standardizes operator libraries and runtime environments and provides an unified PyTorch-compatible programming interface across heterogeneous chips. DiComm implements efficient RDMA communication among diverse chips through a consistent communication API. This design simplifies the integration of heterogeneous hardware into PyTorch-based frameworks, allowing algorithm developers to focus on higher-level optimizations rather than low-level hardware intricacies.

\subsection{DiTorch}

DiTorch is built upon two key principles. First, we employ PyTorch \cite{Ansel_PyTorch_2_Faster_2024} as the unified programming interface layer, a decision justified by its widespread adoption as the de facto standard—owing to its ease of use, flexibility, and high efficiency. Given that PyTorch underpins numerous distributed training systems, such as Megatron\cite{shoeybi2019megatron}, DeepSpeed\cite{rasley2020deepspeed} and InternLM\cite{2023internlm}, it serves as an ideal foundation for ensuring consistency across heterogeneous hardware platforms. Second, acknowledging that large-scale model computations are fundamentally comprised of numerous operators, we have developed a bottom-up precision alignment pipeline tailored for heterogeneous chips. This approach first enforces numerical consistency at the operator level across different chips and then achieves an end-to-end, model-level precision alignment, thereby ensuring the correctness and stability of training large models in hyper-heterogeneous environments.

\subsubsection{Torch Adaptation}

Contemporary training frameworks predominantly rely on PyTorch; however, the runtime interfaces and operator libraries available for heterogeneous chips differ considerably. To address these disparities, DiTorch provides a unified programming interface, thereby simplifying the integration of multiple hardware operator libraries. In practice, merely appending a single line of code enables users to utilize heterogeneous chips in a manner fully consistent with the standard PyTorch distribution.


\begin{tcolorbox}[colframe=blue!50!black, colback=white, title=DiTorch Example for Heterogeneous Hardware, fonttitle=\small]
\begin{verbatim}
>>> import torch
>>> import ditorch
>>> #Unified device name & dispatch key
>>> x = torch.randn(4,4,device="device")
>>> #Precision-verified operator on all chips
>>> y = x + x
\end{verbatim}
\end{tcolorbox}

DiTorch incorporates two major strategies to enhance interoperability and efficiency across diverse AI hardware platforms. Initially, the Torch Adapter may be supplied by the chip vendor if available; otherwise, one can directly employ the \textit{\textbf{Device-Independent Process Unit (DIPU)}}, a Torch Adapter developed within DiTorch. DIPU harmonizes the underlying runtime systems from various vendors into unfied APIs, covering functionalities such as stream and memory management, device control, and kernel launching, thus achieving semantic uniformity for auxiliary (non-operator) operations at the Torch layer. Besides, the \textit{\textbf{Device-Independent Operator Interface (DIOPI)}}\footnote{https://github.com/DeepLink-org/DIOPI} is utilized to bridge the unified Torch operator API with vendor-tailored operator libraries, ensuring consistent semantic interpretation across operators. By establishing a formal computational standard, DIOPI decouples the upper framework layers from the underlying hardware, allowing independent evolution in both realms and promoting the reuse of operators across different systems. Currently, over 300 standardized operator interfaces have been implemented in DIOPI.

\subsubsection{Precision Alignment and Performance Analysis}

DiTorch is furnished with a comprehensive set of operator analysis tools aimed at ensuring numerical accuracy and optimizing performance on distinct AI chip architectures. The suite encompasses modules for both offline and real-time precision evaluation, profilers for execution time analysis and optimization, and mechanisms designed to detect overflow issues in individual or all operators. In addition, DiTorch integrates input-output capture capabilities that extract authentic training data, greatly facilitating debugging and tuning processes. These tools are essential for achieving high-caliber training outcomes for large language models on the supported AI platforms.

\begin{figure}
    \centering
    \includegraphics[width=0.9\linewidth,height=6cm]{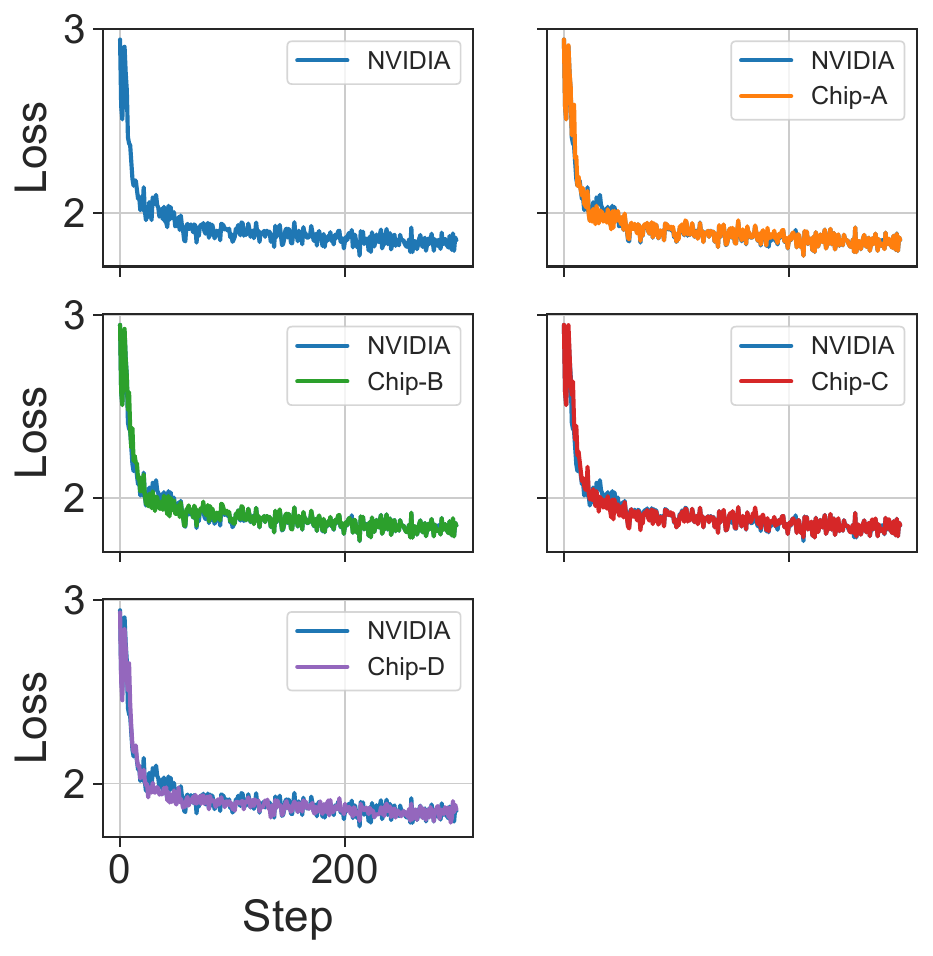}
    \caption{DiTorch's precision alignment across Chips A, B, C, and D compared to the NVIDIA A100.}
    \label{fig:losses_align}
\end{figure}

By leveraging these instruments, DiTorch achieves precision congruity on Chips A, B, C, and D relative to the NVIDIA A100 during large language model training. As shown in Figure \ref{fig:losses_align}, we conducted training of a 20-billion parameter model for 300 iterations using the same dataset. The precision alignment criterion is satisfied if the Mean Relative Error (MRE) remains below 1.5\%:
\[
\frac{1}{n} \sum_{i=1}^{n} \left| \frac{y_i - \hat{y}_i}{y_i} \right| < 1.5\%,
\]
where \( y_i \) represents the loss measured at iteration \( i \) on the NVIDIA A100 GPU, and \( \hat{y}_i \) corresponds to the loss on the evaluated AI chip. In our experiments, we set \( n=300 \). Table \ref{tab:chip_precision} lists the observed MRE values for Chips A, B, C, and D.

\begin{table}
    \centering
    \caption{Observed Mean Relative Error of training loss for the assessed AI chips.}
    \begin{tabular}{@{}c@{}cccc@{}}
        \toprule
          & Chip-A & Chip-B & Chip-C & Chip-D \\
        \midrule
        \multicolumn{1}{@{}c}{MRE} & $0.391\%$  & $0.477\%$  & $0.584\%$  & $1.215\%$  \\
        \bottomrule
    \end{tabular}
    \label{tab:chip_precision}
\end{table}

\subsection{DiComm}

\begin{figure}
    \centering
    \includegraphics[width=0.9\linewidth,height=6cm]{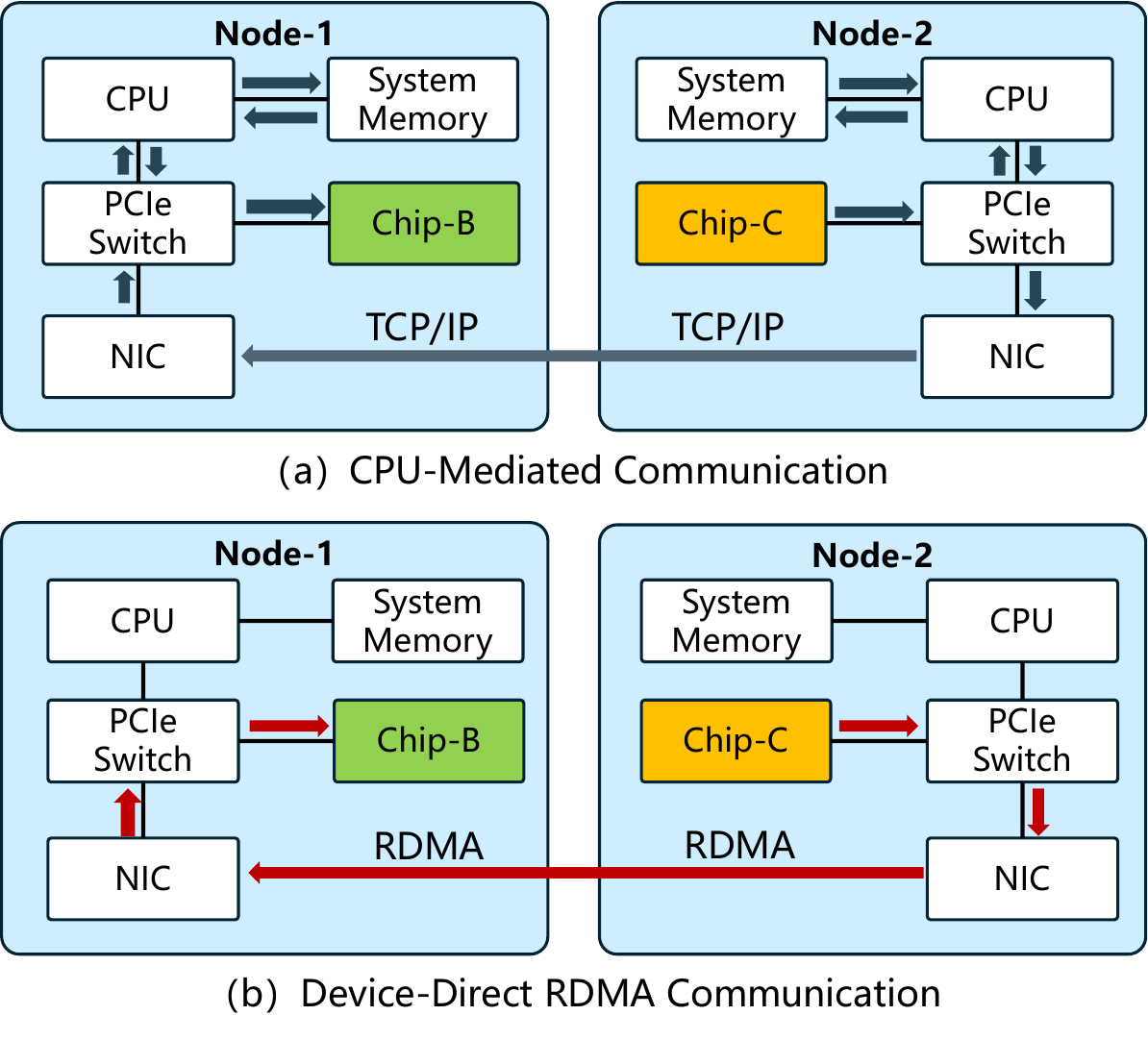}
    \caption{Comparison of CPU-mediated and device-direct communication methods.}
    \label{fig:dicommexample}
\end{figure}

In this work, we introduce DiComm, a unified communication library that overcomes the isolation and inefficiency challenges in complex heterogeneous environments. Leveraging the libibverbs library, DiComm facilitates RDMA communications across various chip architectures and supports both homogeneous and heterogeneous chip-to-chip interactions. DiComm is compatible with protocols including TCP and RDMA, and implements two distinct communication paradigms: CPU-mediated and device-direct. As illustrated in Figure \ref{fig:dicommexample}, we provide detailed comparisons of these protocols and their underlying techniques. Although its primary focus is high-performance peer-to-peer communication, DiComm also incorporates collective communication primitives (e.g., allreduce and broadcast) via a combination of send/receive operations and native communication operators.

\paragraph{CPU-mediated Strategy.}
As depicted in Figure \ref{fig:dicommexample}, the CPU-mediated mechanism in DiComm initiates by transferring data from the source AI chip to the host’s main memory. Thereafter, a CPU-focused communication framework—such as Gloo—is employed to relay this information to the destination node via TCP/IP or RDMA protocols. Following temporary storage in the host memory, the data is subsequently forwarded to the target AI chip. This operational structure ensures wide-ranging compatibility with diverse AI accelerators, making it both adaptable and broadly applicable.

\paragraph{Device-direct Strategy.}
In the device-direct method, each chip registers its local memory regions with an RDMA driver, thereby enabling an RDMA-enabled NIC to map physical addresses for remote access. An RDMA connection manager (e.g., \texttt{rdma\_cm}) then coordinates the connection establishment by configuring queue pairs and exchanging memory region descriptors that incorporate the requisite keys and addresses. After the connection is set up, the NICs utilize their DMA controllers to execute direct read and write operations among device memories, thus bypassing the CPU and host memory entirely. This approach effectively shortens the data transmission path, curtails latency, and enhances overall communication efficiency. To further boost DiComm’s performance, we deliberately select NICs that are particularly well-suited for inter-chip communications.

\subsubsection{DiComm Performance} 

\begin{figure}
    \centering
    \includegraphics[width=0.9\linewidth]{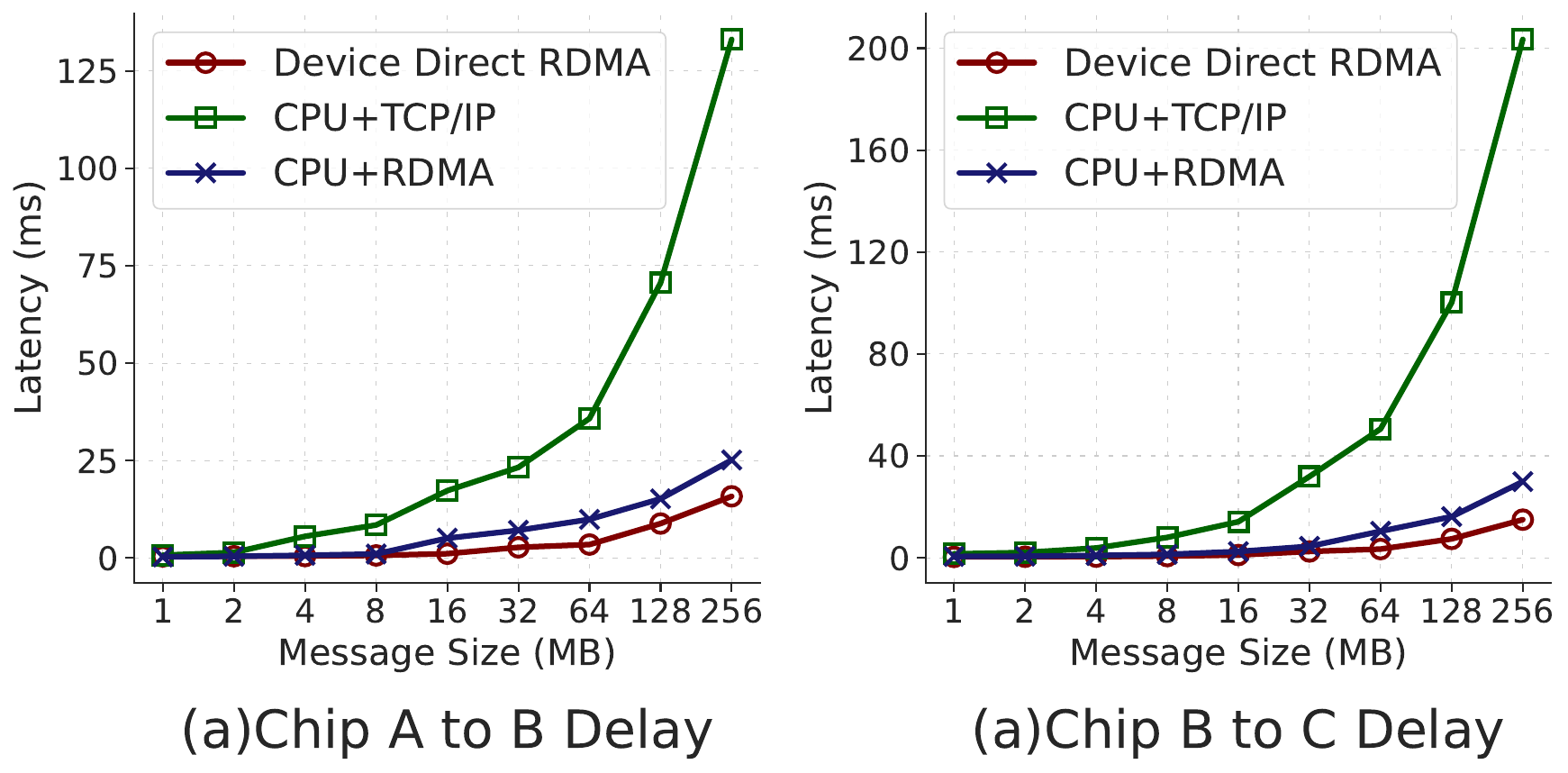}
    \caption{Comparison of cross Chip-Communication latency with different strategies.}
    \label{fig:Latency_Comparison}
\end{figure}

DiComm successfully enables P2P communication between different chips using the device-direct strategy. As shown in Figure~\ref{fig:Latency_Comparison}, device-direct RDMA reduces average latency by 9.94× compared to the conventional TCP/IP scheme, with speedups ranging from 1.79× to 16.0× depending on the message size.

Given that the GLOO backend in the PyTorch framework currently lacks support for RDMA and is limited to TCP, DiComm offers substantial performance enhancements for heterogeneous interconnects over the PyTorch GLOO approach.

\section{Heterogeneous Parallelism}

In this section, we delineate the key observations that underpin the design of \textbf{\textit{HeteroPP}}, a large-scale training algorithms for LLMs on heterogeneous chip architectures. Our analysis is motivated by three dimensions of hardware discrepancy—memory capacity, communication bandwidth, and computational power—coupled with the unbalanced distribution of chip counts and the expansive parallel and model scales intrinsic to modern language model training.

\subsection{Key Observations for HeteroPP}

In traditional homogeneous distributed training, we have the following observations and ideas how they can be applied to heterogeneous training: 

\textbf{Observation \#1:} \textit{Both data parallelism and tensor parallelism require high-performance AllReduce and some other collective communication operators. In contrast, the point-to-point communication inherent in pipeline parallelism can be more readily overlapped with computation.}


In actual heterogeneous cluster environments, different heterogeneous chips are distributed across separate server nodes, resulting in inherently homogeneous intra-node configurations. Moreover, in our ultra-heterogeneous scenario, the underlying software stacks and hardware topologies vary for each type of chip, making the development of a unified high-performance collective communication operator extremely challenging. As the diversity of chip types increases, so do the engineering workload and complexity. Therefore, leveraging pipelined parallelism to connect different heterogeneous chips emerges as the most rational and efficient approach.

\textbf{Observation \#2:} \textit{Tensor parallelism is extremely sensitive to bandwidth and typically occurs within nodes with high bandwidth.}

Due to the varying internal designs and hardware topologies of heterogeneous chip nodes, there are differences not only in the number of chips but also in their intra-node interconnect configurations. In certain chip server environments, noticeable throughput degradation is observed when intra-node communication involves traversing NUMA boundaries or crossing PCIe switches. This variability in interconnect performance consequently influences the optimal range for configuring tensor parallelism dimensions.

\textbf{Observation \#3:} \textit{The efficiency of pipeline parallelism is closely related to the load balance of the tasks assigned to each pipeline stage.}

In homogeneous distributed training, pipeline load balancing is achieved through uniform layer sharding across each stage and consistent parallel dimensions. In heterogeneous mixed training, however, the computational and communication capabilities vary across different chip types. Consequently, non-uniform layer sharding and stage-specific adjustments of parallel dimensions become highly valuable. Additionally, configuration complexity can be reduced by applying identical settings to homogeneous chips while using distinct settings for heterogeneous chips.

\textbf{Observation \#4:} \textit{Within the pipeline parallelism, the early stages require a larger number of warmup micro-batches to ensure seamless computational flow to subsequent stages. This necessity imposes additional memory demands.}

Due to differences in memory specifications among heterogeneous chips, we can allocate chips with larger memory to the earlier stages of pipeline parallelism, while those with smaller memory can be assigned to the later stages. This not only reduces the need for recomputation but also expands the available parallel dimensional space for each chip type.

\subsection{HeteroPP Design}

\begin{figure}
    \centering
    \includegraphics[width=1.0\linewidth, height=9cm]{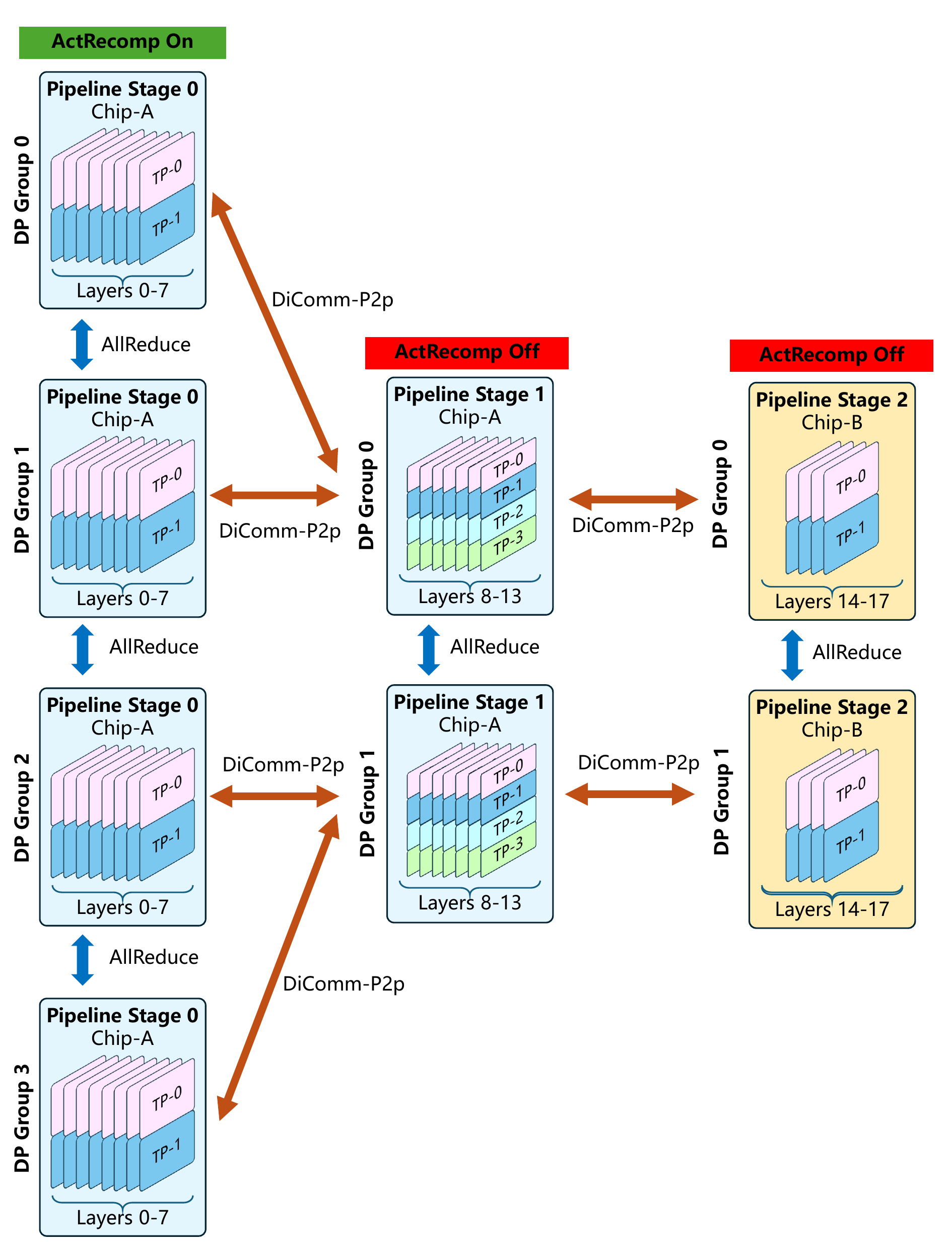}
    \caption{A simple example of HeteroPP. In this setup, two types of chips are used, and the model consisting of 18 layers is partitioned into three pipeline stages. At each pipeline stage, different configurations for data parallelism, tensor parallel dimensions, layer sharding and activation recomputation settings can be employed.}    
    \label{fig:heteropp_arch}
\end{figure}


In this work, we extend pipeline parallelism to heterogeneous AI Chip-Clusters for distributed LLM training. Figure \ref{fig:heteropp_arch} illustrates a simple example of the HeteroPP framework. In this example, two types of chips,A and B, are used to train an 18-layer model, with 16 chips A and 4 chips B. Each pipeline stage is composed entirely of a single chip type, and each chip type can be mapped to multiple pipeline stages. In our example, chip A occupies the first two stages, and chip B occupies the final stage. The mapping order corresponds to the chips’ memory capacities in descending order. Within each pipeline stage, homogeneous chips are partitioned into multiple TP and DP groups, where TP sizes 2,4,2 and DP sizes 4, 2, 2 are used in three pipeline stages separately. Activation recomputation is enabled for the first pipeline stage to relieve memory stress and disabled for the last two pipeline stages to increase computation efficiency. 8, 6, and 4 layers are shared into three pipeline stages separately. P2P communication is facilitated by DiComm, which supports both a CPU-mediated strategy and a device-direct strategy. Gradient synchronization within DP groups is performed using AllReduce operations, constrained to AI chips of the same type. 

As we can see, compared to homogeneous pipeline parallelism, HeteroPP introduces several distinct characteristics due to computational and memory capacity imbalances among different AI chips.

Based on Observation \#1, the HeteroPP framework is architected so that each pipeline stage consists exclusively of homogeneous chip nodes, with heterogeneous nodes being strategically distributed across different stages. Additionally, drawing from Observation \#4, we \textbf{map chips with larger memory capacities to the earlier stages of the pipeline}, while those with smaller memory are allocated to the later stages. This allocation strategy not only optimizes resource utilization but also enhances communication efficiency throughout the pipeline.

Moreover, to achieve balanced pipeline workloads due to the Observation \#3 and fully exploit the computing, communication, and memory capabilities of the chips, our system supports non-uniform task distribution in three key aspects. First, we enable \textbf{uneven partitioning of model layers}. This approach allows for the allocation of a greater number of layers to pipeline stages with stronger computational resources or to heterogeneous chips with larger memory capacities. Second, we allow for \textbf{flexible adjustments in the intra-stage parallel dimensions}, encompassing both data-parallel and tensor-parallel dimensions. This flexibility enables dynamic tuning of the number of chips per stage, as the total chip count is determined by the product of the data-parallel and tensor-parallel dimensions.  Furthermore, \textbf{activation recomputation can be flexibly enabled or disabled} based on the memory requirements of each pipeline stage and the available chip memory.


Despite these differences, HeteroPP remains compatible with a variety of pipeline parallelism scheduling strategies, including Chimera \cite{Chimera}, ZB-V \cite{qi2024zero} and ZeroPP \cite{tang2024zeroppunleashingexceptionalparallelism}.

\subsection{Automatic Search for HeteroPP Strategy}

\begin{figure}
    \centering
    \includegraphics[width=1.0\linewidth, height=10cm]{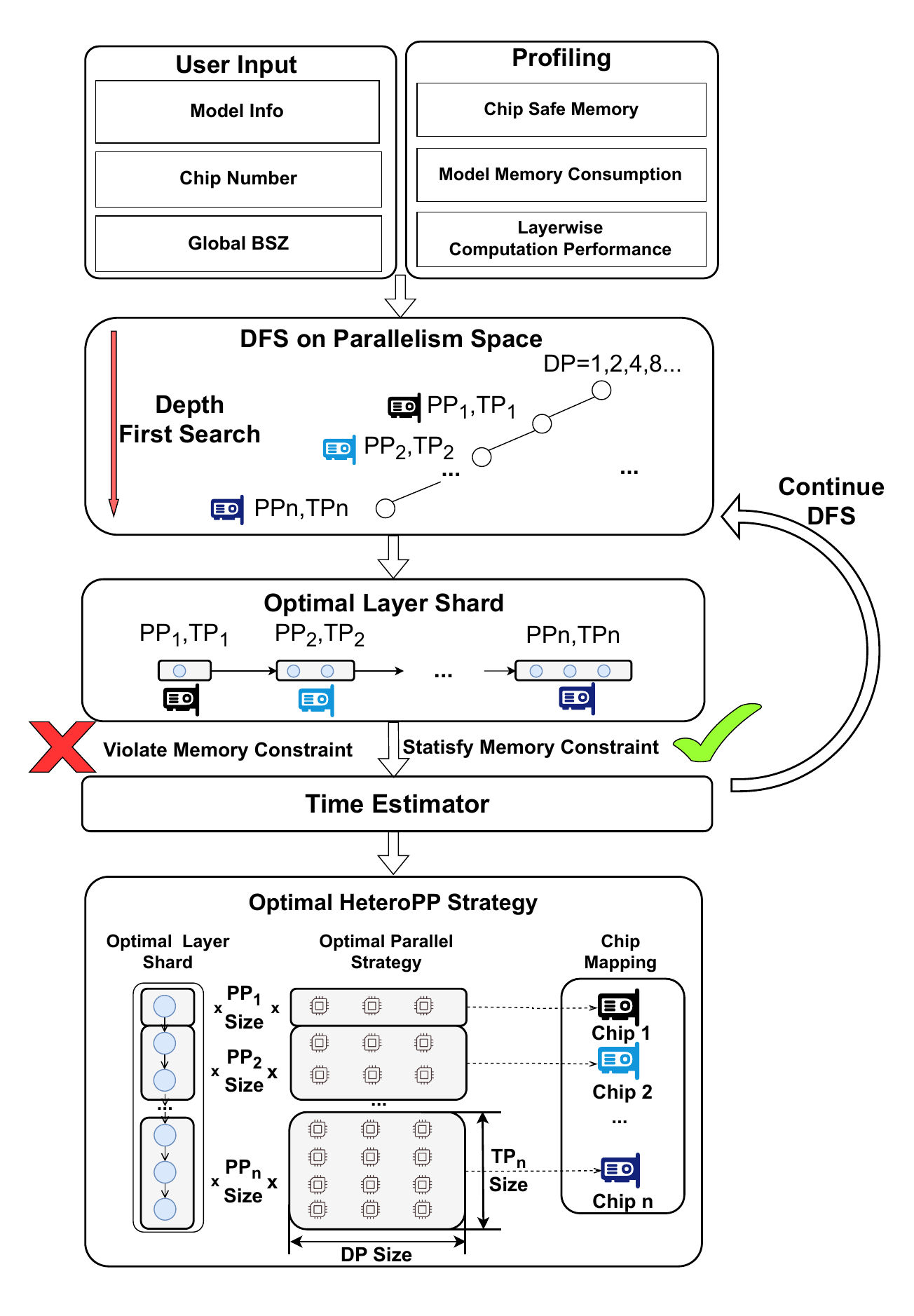}
    \caption{Overview of HeteroAuto.}    
    \label{fig:heteroauto_arch}
\end{figure}

We introduce \textbf{\textit{HeteroAuto}}, an automated search method for optimizing parallel strategy configurations in heterogeneous chip environments within the HeteroPP framework. Our approach tackles the dual challenges of \textbf{resource heterogeneity} and \textbf{communication overhead}. Specifically, HeteroAuto is designed to identify efficient heterogeneous training parallelization configurations for extremely large models on hyper-heterogeneous clusters. At the same time, we ensure that the search time remains within an acceptable range, balancing optimal performance with practical usability.

Besides the observations listed in previous section, our design is motivated by an extra factors: in training extremely large models, memory constraints often restrict the micro-batch size to 1, thereby limiting the potential for adjustments solely via data parallelism.


\subsubsection{Search Space.} 
Given an LLM with \( L \) Transformer layers, we train the model on a heterogeneous cluster comprising \( \sum_{i=1}^{C} N_i \) AI chips, where \( C \) represents the number of different chip types, and \( N_i \) denotes the number of chips of type \( i \). The global batch size is fixed at \( B \).  
HeteroAuto explores the following decision variables to optimize overall training performance. It searches for \( s_{pp,i} \), the number of pipeline stages assigned to chip type \( i \) determining the total pipeline parallelism degree as \(  s_{pp} = \sum_{i=1}^{C} s_{pp,i} \), and \( s_{tp,i} \) which represents the tensor parallelism degree used by chip type \( i \) during training.

Given a total of \( L \) layers, HeteroAuto also determines the layer allocation strategy by searching for \( l_i \), the number of layers assigned to chip type \( i \). Consequently, the number of layers per pipeline stage on chip type \( i \) is given by \( l_i / s_{pp,i} \). In this formulation, based on Observation \#3, we assume that layers are \textbf{\textit{non-uniformly}}  distributed across different chip types and \textbf{\textit{evenly}}  distributed within the same chip type. 

Additionally, HeteroAuto independently determines the recompute setting for each chip type. We denote this setting by \(r_i\), where \(r_i\) can take the value of 0 or 1, corresponding to recompute being disabled or enabled, respectively.
Furthermore, HeteroAuto determines \( s_{dp} \), the data parallelism degree, which in turn defines the number of micro-batches used in pipeline parallelism as \( b = B / s_{dp} \).

\begin{table}
    \centering
    \caption{Notations used in HeteroAuto.}
    \begin{tabular}{@{}l@{}l@{}}
        \toprule
        \textbf{Notation } & \textbf{Description} \\  
        \midrule
        \( C \) & Number of different chip types in the cluster. \\  
        \( N_i \) & Number of chips of type \( i \), $N_i = s_{pp,i} \times s_{tp,i} \times s_{dp}$. \\  
        \( B \) & Global batch size. \\  
        \( L \) & Total number of Transformer layers in the model. \\  
        \( \alpha \) & Bubble coefficient for pipeline schedule. \\
        \( s_{dp} \) & Number of data parallelism degree.  \\  
        \( b \) & Number of micro-batch, $b = B/s_{dp}$. \\  
        \( s_{pp,i} \) & Number of pipeline stages allocated to chip type \( i \). \\  
        \( s_{pp} \) & Pipeline parallelism degree, \( S_{pp} = \sum_{i=0}^{C-1} s_{pp,i} \). \\  
        \( s_{tp,i} \) & Tensor parallelism degree used by chip type \( i \). \\  
        \( l_i \) & Number of layers assigned to chip type \( i \). \\  
        \( r_i \) & Recomputation setting of chip type \( i \). \\
        \bottomrule
    \end{tabular}
    \label{tab:notation}
\end{table}


\subsubsection{Cost Model.} 


HeteroPP seeks an optimal solution that minimizes the estimated iteration time:
\[
T = \min \left\{ \max_{1 \leq i \leq s_{pp}} \big( b \cdot T_i^{\text{comp}} + T_i^{\text{update}} + \alpha \cdot \sum_{j=1,j\neq i}^{s_{pp}} T_j^{\text{comp}} \big) \right\},
\]
where \( T_i^{\text{comp}} \) denotes the per-pipeline-stage computation time of chip \(i\) for a single microbatch, including both forward, backward computation as well as recomputation overhead. Thus, \( T_i^{\text{comp}} \) can be further refined and expressed by the following equation, where \( t_{s_{tp,i}}^{fwd} \), \( t_{s_{tp,i}}^{bwd} \) and  \(t_{s_{tp,i}}^{recomp}\) represent layer-wise forward, backward and recomputation time with TP size \(s_{tp,i}\) on chip \(i\):

\[
T_i^{\text{comp}} = ceil(\frac{l_i}{s_{pp,i}}) \cdot (t_{s_{tp,i}}^{fwd} + t_{s_{tp,i}}^{bwd} + r_i \cdot t_{s_{tp,i}}^{recomp})
\]

\( T_i^{\text{update}} \) represents the per-pipeline-stage optimizer update time of chip \(i\), which consists of the optimizer computation time and the gradient synchronization time that is not overlapped by the backward computation. \( T_i^{\text{update}} \) can be further refined and expressed by the following equation, where \(t_{s_{dp}, s_{tp,i}}^{update}\) represents layer-wise update time of chip $i$ given tp size \(s_{tp,i}\) and dp size \(s_{dp}\):

\[
T_i^{\text{update}} = ceil(\frac{l_i}{s_{pp,i}}) \cdot t_{s_{dp}, s_{tp,i}}^{update}
\]

In addition, there exist numerous pipeline parallelism strategies, each with distinct methods for computing the bubble time. Under large-scale training scenarios, the bubble time can be considered proportional to the computation time of a single microbatch. Therefore, we introduce a configurable bubble coefficient \( \alpha \). For instance, in zero-bubble pipeline parallelism approaches such as ZB-V\cite{qi2024zero}, \( \alpha \) is set to 0. In our work, since we employ the conventional 1F1B approach, we set \( \alpha \) to be 1, and the computation formula for \(T\) the computation formula is equivalent to the one used in Metis\cite{um2024metis}. Therefore, \( T \) represents the total iteration time, which accounts for the maximum computation time across all pipeline stages, the optimizer step time, and the pipeline bubble overhead.  

We use an auto-profiler to profile the layer-wise performance of each chip, including \( t_{s_{tp,i}}^{fwd} \), \( t_{s_{tp,i}}^{bwd} \),  \(t_{s_{tp,i}}^{recomp}\) and \(t_{s_{dp}, s_{tp,i}}^{update}\) under various DP and TP sizes. Additionally, we will profile layer-wise memory consumption with and without activation recomputation.


Based on our empirical findings, we enforce the following requirements in our search for heterogeneous parallel strategies:
\begin{enumerate}
\setlength{\itemsep}{0pt}  
\setlength{\parskip}{0pt}
    \item For pipeline stages associated with the same chip category, a uniform tensor-parallel degree must be employed, and each stage should be allocated an identical number of Transformer layers.
    \item For every chip indexed by \(i \in [1,C]\), the tensor-parallel degree \(s_{tp,i}\) is constrained to powers of two (e.g., 1, 2, 4, 8, etc.). Additionally, to optimize communication efficiency, \(s_{tp,i}\) is generally capped at \(TP\_MAX_i\), a limit determined by the number of chips available within a single node, NUMA domain, or PCIe switch depending on the hardware configuration.
    \item The overall memory usage must remain within the safe capacity profiled for each individual chip.
\end{enumerate}


\subsubsection{Search Algorithm.}

To thoroughly explore the vast decision space, we adopt a depth-first search (DFS) methodology outlined by the following procedures:  
\begin{enumerate}
\setlength{\itemsep}{0pt}  
\setlength{\parskip}{0pt} 
    \item \textbf{Depth-First Search for Parallelism Space.}  
    Select candidate values for data parallelism from the allowed set, ensuring that each candidate evenly divides the global batch size $B$. For every chosen value $s_{dp}$, compute the associated micro-batch count as $b = B/s_{dp}$. For each candidate $s_{dp}$ and for every chip type $i$, choose a tensor parallelism size \(s_{tp,i}\) from the set $\{1, 2, \ldots, \\ TP\_MAX_i\}$, and then determine the corresponding pipeline parallelism size \(s_{pp,i}\) using the relation $N_i = s_{pp,i} \times s_{tp,i} \times s_{dp}$. A DFS procedure then systematically explores these options across different chip types, ordered by descending memory capacity.
    
    \item \textbf{Optimal Layer Sharding.}  
    For every configuration represented as $(s_{dp}, b, \{(s_{pp,i}, s_{tp,i}, r_i) \mid i \in [1,C]\})$, perform a heuristic search to derive the best layer distribution. An initial assignment is computed by equalizing the compute time among chip groups. If this allocation exceeds the total number of layers $L$, an iterative refinement readjusts the per-group assignments while ensuring that memory usage remains within safe limits.
    
    \item \textbf{Cost Estimation and Strategy Selection.}  
    Among the configurations that meet memory constraints, estimate the execution time per iteration. The computation time for processing a single layer with one micro-batch on the $i$-th chip (given a specific tensor parallel configuration) is pre-profiled. The configuration that minimizes the total iteration time is then adopted as the optimal parallel strategy.
\end{enumerate}  

A \textbf{two-stage} procedure further refines the search for parallelism strategies, enhancing output efficiency with only a minimal increase in search overhead. In the first stage, HeteroAuto identifies a suitable data parallelism size exclusively. In the subsequent stage, a sizable group of homogeneous chips is partitioned into smaller subgroups, each treated as a distinct heterogeneous entity. These groups, combined with the DP size established earlier, are input into the algorithm once again. Additionally, an extra constraint can be enforced among groups of the same chip type: if group $a$ precedes group $b$, then it must hold that $s_{tp,a} \ge s_{tp,b}$, thereby pruning less promising candidates. This strategy allows for a more nuanced exploration of heterogeneous parallelism, ultimately leading to enhanced efficiency in parallel training.
\section{Optimization}\label{opt}

\begin{figure}
    \centering
    \includegraphics[width=0.95\linewidth,height=6cm]{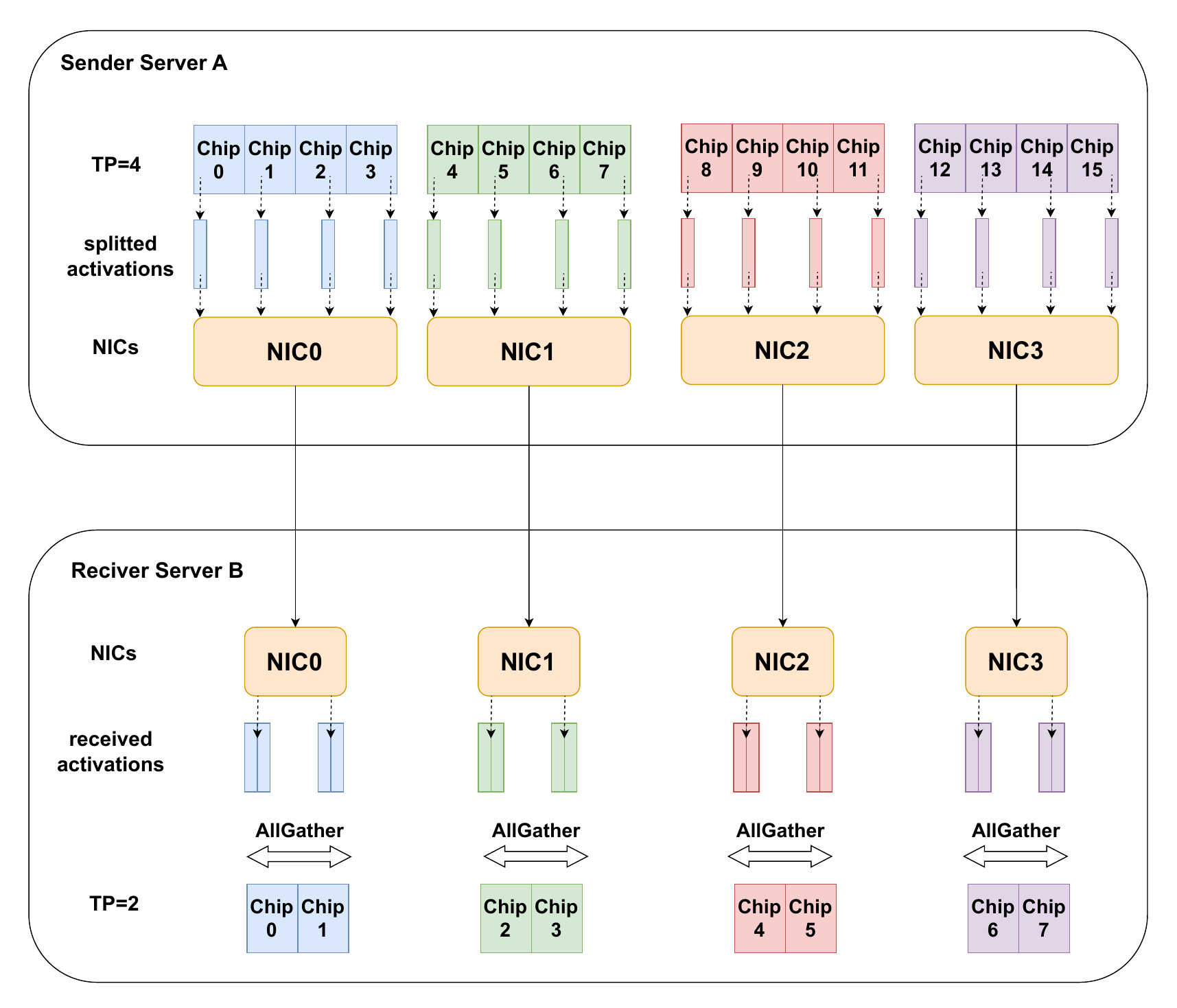}
    \caption{An example of topology-aware send/recv + all-gather activation resharding with TP size 4 on Chip-A server and TP size 2 on Chip-B server.}    
    \label{fig:topoaware}
\end{figure}

To enhance training efficiency in heterogeneous settings, we study activation resharding in parallel architectures. Here, activation resharding denotes transferring activation values between consecutive pipeline stages which can be mapped to different types of chips severs.

Under the assumption that each GPU server is equipped with only a single NIC and that the NIC's bandwidth is the primary bottleneck for cross-node activation resharding, previous work~\cite{zhuang2023optimizing} has proposed a broadcast-based method for inter-node activation remapping. This approach minimizes the volume of data transferred between nodes and allows for overlapping data transfers among different GPUs within the same node. However, hyper-heterogeneous environments face additional challenges:
    
\begin{enumerate}
    \item GPU servers of different chips have multiple NICs with varying counts and affinities.
    \item PCIe links between PCIe switches and chips can limit bandwidth, requiring concurrent transmissions of multiple chips to saturate a single NIC.
\end{enumerate}

To address these issues, we propose a topology-aware activation resharding strategy as shown in Figure~\ref{fig:topoaware}. 

First, we configure each chip server to assign a dedicated communication NIC to each chip based on its NIC affinity. This setup not only balances the load across the NICs but also minimizes the overall communication path length. The performance enhancement achieved by utilizing affinity NICs is evident in Table \ref{tab:nic_affinity_comparision}.

\begin{table}
\centering
\caption{Throughput comparison between NIC affinity and non-affinity on two heterogeneous servers using 8 chips for concurrent communication with message size 64MB.}
\resizebox{1.0\linewidth}{!}{%
\begin{tabular}{|c|c|c|c|}
\hline
Chips & Non-affinity NIC& Affinity NIC  & Improvement\\ 
&  (GB/s) & (GB/s) & \\ \hline
Chip A $\rightarrow$ B & 5.51 $\times$8 & 9.56 $\times$8 & 73.5\% \\ \hline
Chip B $\rightarrow$ D & 5.23 $\times$8 & 9.91 $\times$8 & 89.5\% \\ \hline
\end{tabular}%
}
\label{tab:nic_affinity_comparision}
\end{table}

Second, we employ a communication mode that combines send/recv with an all-gather operation. This strategy optimizes the system by minimizing cross-node data transfers. Additionally, since multiple chips share a single NIC during communication, the available bandwidth on each NIC is maximized. Given that our tensor parallel communications predominantly occur within the node, the communication overhead incurred by the all-gather operation within the node is negligible compared to inter-node communication delays.


To further reduce communication overhead in pipeline parallelism, we have implemented a fine-grained overlap between point-to-point (P2P) communication and computation at the framework level. Drawing inspiration from previous works~\cite{qi2024zero,chen2024optimizing}, we decompose the traditional forward-backward computation into four distinct phases: forward computation, backward recomputation, backward input gradient calculation, and backward weight gradient calculation. This decomposition enables a more precise interleaving of computation with P2P communication. Consequently, our approach achieves near-lossless P2P communication performance, substantially minimizing the communication overhead in pipeline-parallel training.
\section{Main Evaluations}

\subsection{Experimental Setup}

\subsubsection{Model Configuration}

\begin{table}
    \centering
    \caption{Model configuration used in this work.}
    \begin{tabular}{@{}llll@{}}
        \toprule
        \textbf{Parameter} & \textbf{Value} & \textbf{Parameter} & \textbf{Value} \\ \midrule
        \# Layers & 96 & Hidden Size & 8192 \\
        \# Attention Heads & 64 & Intermediate Size & 36864 \\
        \# Queries per Head & 8 & Vocabulary Size & 92544 \\
        Max Sequence Length & 4096 &  &  \\
        \bottomrule
    \end{tabular}
    \label{tab:modelconfig}
\end{table}
The architectural configuration of the 100B-parameter model\cite{cai2024internlm2}, as presented in Table~\ref{tab:modelconfig}, adheres to the structural design principles established by LLaMA~\cite{grattafiori2024llama3}. To further enhance memory efficiency during inference, the model incorporates Group Query Attention, a technique that effectively reduces memory usage while maintaining performance.

\subsubsection{Chip Specification}

\begin{table}
    \centering
    \caption{Performance comparison of AI chips relative to the A100 used in this study.}
    \begin{tabular}{llll}
        \toprule
        \textbf{Chip} & \textbf{FP16} & \textbf{Memory} & \textbf{\#Chips}   \\
         &\textbf{(TFLOPS)} & \textbf{(GB)} & \textbf{per Node} \\
        \midrule
        A & >$0.5$, <$1.0 \times \text{A100}$ & 96 & 16  \\
        B & >$0.5$, <$1.0 \times \text{A100}$ & 64 & 8 \\
        C & >$0.0$, <$0.5 \times \text{A100}$ & 32  & 16 \\
        D & >$1.5$, <$2.0 \times \text{A100}$ & 32 & 8 \\
        \bottomrule
    \end{tabular}
\label{tab:chip_comparison}
\end{table}

We tested our framework with four different types of AI chips. A detailed comparison of the AI accelerators used in this study is presented in Table \ref{tab:chip_comparison}. Notably, Chip-D outperform the NVIDIA A100 GPU in computational power, although they are constrained by limited memory capacity. Among the accelerators deployed, only Chip-A offers a larger memory capacity than the NVIDIA A100 GPU.

Figure \ref{tab:homo3dconfig} presents the training throughput of the 100B-parameter model, measured in tokens per chip per second (TGS\footnote{TGS is typically used to denote tokens per GPU per second; however, we use it here for consistency with previous studies.}), on homogeneous clusters using Chips A, B, C and D.  Tables \ref{tab:homo3dconfig} shows the used hybrid parallelism configurations. Each chip's training efficiency is evaluated on a 256-chip setup with a global batch size of 2M tokens.  Chip-B achieves the highest throughput at 143.7 TGS. In contrast, Chip-C demonstrates the lowest performance, reaching only 46.2 TGS, primarily due to its lower FP16 TFLOPS, as detailed in Table \ref{tab:chip_comparison}. Interestingly, despite Chip-D having the highest theoretical peak performance, its actual throughput is limited to 99.5 TGS due to memory constraints and communication bottlenecks. To mitigate these limitations, we employ CPU offloading and tensor parallelism for Chip-D, though this significantly impacts its overall training efficiency.


\begin{table}[t]
    \centering
    \caption{Training throughput (tokens per chip per second, TGS) comparison and hybrid parallelism configurations for homogeneous training among Chips A, B, C and D on 256 chips.}
    \begin{tabular}{ccccccc}
        \toprule
        Chip  & PP & DP & TP & Extra & TGS\\
        \midrule
        Chip-A   & 16 & 4 & 4 & N/A & 136.9 \\
        Chip-B   & 16 & 4 & 4 & Activation Recompute & 143.7 \\
        Chip-C   & 32 & 2 & 4 & Activation Recompute & 46.2 \\
        Chip-D   & 8 & 4 & 8 & CPU Offload & 99.5 \\
        \bottomrule
    \end{tabular}
    \label{tab:homo3dconfig}
\end{table}

\subsubsection{Cluster Information}
In this study, all AI chips within each of the nine clusters are interconnected via a multi-rail RDMA network utilizing RoCE-v2. The network is configured with a 1:1 oversubscription ratio to ensure optimal data transmission efficiency.

\subsection{Evaluations on HeteroPP and HeteroAuto}
To validate the performance and efficiency of the proposed heterogeneous parallelism strategy, we conducted a series of extensive experiments focusing on HeteroPP and HeteroAuto. Various chip configurations, along with their corresponding global batch sizes (GBS), were evaluated. These configurations are summarized in Table~\ref{tab:heteroexpconfig}.

We use the \textit{HeteroSpeedupRatio} to evaluate heterogeneous training performance, defined as the ratio of training throughput on the heterogeneous cluster to the sum of the baseline training throughput of each AI chip type in Table \ref{tab:homo3dconfig}:  
\[
HeteroSpeedupRatio=\frac{N \cdot TGS}{\sum_{i=1}^{C}{N_i \cdot TGS_i}}.
\]

\begin{table}
    \centering
    \caption{Chip-Configurations and corresponding global batch sizes for HeteroPP and HeteroAuto.}
    \resizebox{1.0\columnwidth}{!}{%
    \begin{tabular}{ccc}
        \toprule
        Index & Chip-Configuration & GBS (Tokens) \\
        \midrule
        Exp-A-1 & Chip-A (256) + B (256) + C (256)  & 2M \\
        Exp-A-2 & Chip-A (256) + B (256) + C (256)  & 6M \\
        Exp-B-1 & Chip-A (256) + B (256) + C (256) + D (256) & 2M \\
        Exp-B-2 & Chip-A (256) + B (256) + C (256) + D (256) & 8M \\
        Exp-C-1 & Chip-A (384) + B (1024) & 4M \\
        Exp-C-2 & Chip-A (384) + B (1024) & 8M \\
        Exp-D   & Chip-A (384) + B (2048) & 8M \\
        \bottomrule
    \end{tabular}%
    }
    \label{tab:heteroexpconfig}
\end{table}

\begin{figure}
    \centering
    \includegraphics[width=1.0\linewidth,height=4.5cm]{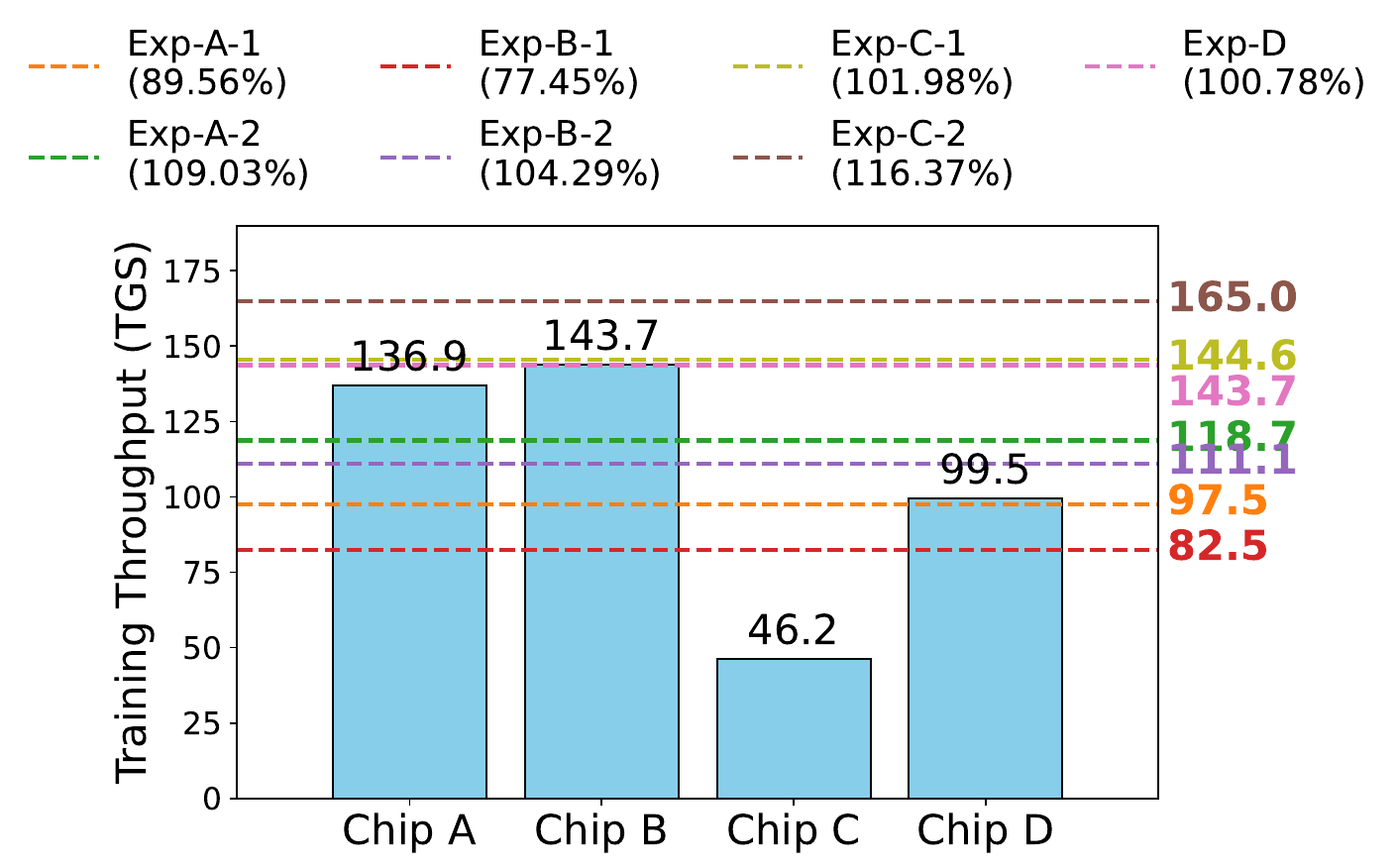}
    \caption{Training throughput for individual homogeneous and heterogeneous training setups, and HeteroSpeedupRatio for heterogeneous training setups.}
    \label{fig:scalability}
\end{figure}

\subsubsection{Effectiveness.}
In the cases of Exp-A to Exp-D, which utilized Chips A, B, C and D, we compared the throughput of each chip type with that achieved through their combined heterogeneous training configurations. In Exp-A and Exp-B, Two heterogeneous training configurations were evaluated. In the first configuration, the global batch size was the same as that used for individual chip-group training. In the second configuration, the global batch size equaled the sum of the batch sizes employed in individual chip-group training. The results, as shown in Figure~\ref{fig:scalability}, demonstrate that the heterogeneous training approach delivers competitive, and in some instances, superior efficiency compared to the baseline throughput of homogeneous training.

\begin{tcolorbox}[colframe=black, colback=white, boxrule=1pt]
For the configuration where GBS is set to the sum, DiTrain achieves, a 109.03\% \textit{HeteroSpeedupRatio} with 768 chips in Experiment A, involving 3 chip types, and 104.29\% with 1,024 chips in Experiment B, incorporating 4 chip types. For constant GBS, \textit{HeteroSpeedupRatio} can still achieve 89.56\% and 77.45\% respectively.
\end{tcolorbox}

Although the observed superlinear performance improvement may appear counterintuitive, it is explainable. The conventional 3D parallel training tends to overlook the imbalanced resource requirements among various computational tasks, while the HeteroPP framework with HeteroAuto capitalizes on these imbalances by intelligently allocating chip tasks and fine-tuning training hyperparameters based on the specific resource demands. Taking Exp-C as example, Chip B has better computational power but is limited by memory, requiring a TP size of at least 8 or activation recomputation enabled, which reduces efficiency due to communication overhead or extra computation tasks. Chip A, while less powerful, has more global memory. In the heterogeneous setup, Chip A handles the earlier pipeline stages, which requires more memory, while Chip B is used for later stages where a smaller TP size is sufficient and activation recomputation can be turned off. This configuration allows Chip B to fully utilize its computational power and chip A to fully utilize its memory capacity, improving overall efficiency. 


In practical production environments, lower-spec chips usually feature significantly lower pricing and reduced power consumption compared to their high-spec chips. By leveraging the HeteroPP framework for heterogeneous training, which integrates both lower-spec and high-spec chips, we can achieve training performance comparable to or even exceeding that of homogeneous high-spec setups, which preserves high performance and reduces overall training costs.

\subsubsection{Searching Overheads.} Beyond performance and scalability, we also evaluated the computational overhead of the strategy search process. Due to the diverse set of heterogeneous chips used in our training, directly comparing with other heterogeneous strategy search algorithms (e.g., Metis\cite{um2024metis} and Alpa\cite{zheng2022alpa}) requires significant adaptation efforts. However, for reference, Metis takes 600 seconds to complete a search for only 64 chips and two chip types, while Alpa takes 240 minutes for the same task. Table~\ref{tab:overhead} summarizes the overheads for our search strategy in three configurations. The search is implemented as a single-threaded Python script running on an Intel® Xeon® Platinum 8460Y+ CPU model, and the two-stage search algorithm described in Section 4 is applied. In the second stage, every set of 128 homogeneous chips is treated as a single group of heterogeneous chips.

\begin{table}
    \centering
    \caption{Overhead of the strategy search process.}
    \begin{tabular}{ccccc}
        \toprule
          & Exp-A & Exp-B & Exp-C \\
        \midrule
         Time (s) & 0.62 & 5.48 & 12.29 \\
        \bottomrule
    \end{tabular}%
    \label{tab:overhead}
\end{table}

\subsection{Ablation Study}

\begin{figure}
    \centering
    \includegraphics[width=1.0\linewidth,height=3.5cm]{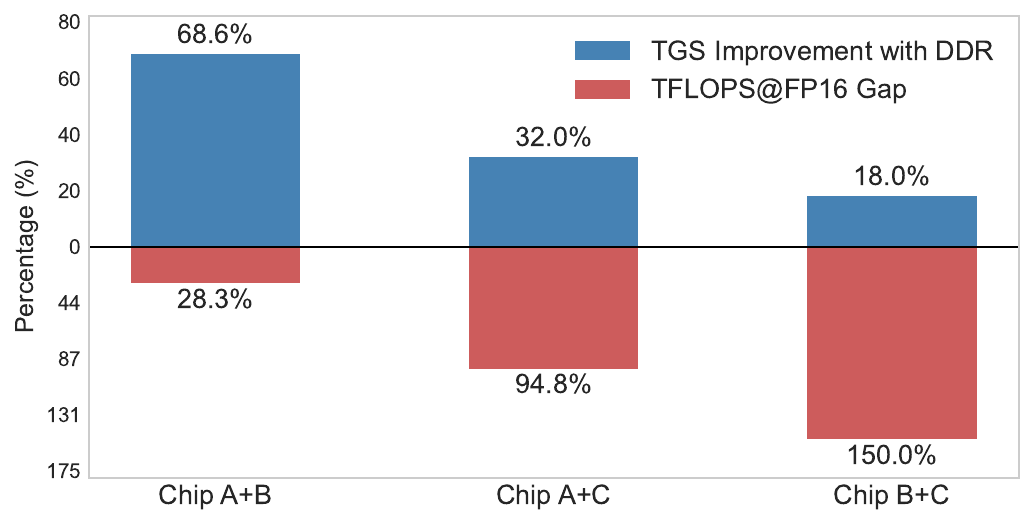}
    \caption{For end-to-end training of a small-scale 8-decoder-layer model, experiments were conducted with and without DDR. Uniform 1F1B schedule was employed, and the parallelism configuration was set to TP=4, PP=2, and DP=2. Two heterogeneous servers, with eight chips on each server, were utilized.}
    \label{fig:end2end_small}
\end{figure}

Figure \ref{fig:Latency_Comparison} provides a detailed comparison of the latency across various message sizes for different communication methods. To more intuitively demonstrate the impact of DiComm, we conducted an end-to-end performance comparison between CPU-mediated TCP and device-direct RDMA communications. This evaluation encompasses both large-scale and small-scale models as well as various chip configurations. The experimental results for small-scale end-to-end training are presented in Figure \ref{fig:end2end_small}.


The results clearly show that device-direct RDMA mode consistently yields substantial performance gains over CPU-mediated TCP across all chip combinations. However, the degree of improvement varies: the performance gap is relatively small between Chip A and B, whereas Chip C exhibits a markedly larger discrepancy compared to the others. Consequently, under the current parallel strategy, Chip C becomes the computational bottleneck, thereby limiting the overall benefits derived from optimizing pipeline parallel P2P communication. These observations suggest that the refinement of heterogeneous parallel strategy is essential to fully take advantages of DiComm and achieve enhanced training efficiency.

In large-scale training scenarios, the increased computational load occupies a larger portion of the overall workload, effectively masking more substantial communication latencies. Consequently, the performance gap between device-direct RDMA and CPU-mediated TCP communication methods narrows, and HeteroPP brings obvious throughput improvement as demonstrated in Table \ref{tab:ablation}.



\begin{table}
    \centering
    \caption{Ablation study variants for large-scale heterogeneous training using Exp-C-1 configuration. SR\&AG resharding refers to the send/recv and all-gather activation resharding mentioned in section \ref{opt}.}
    \resizebox{1.0\linewidth}{!}{%
    \begin{tabular}{|c|c|c|c|c|}
        \hline
         \shortstack{\textbf{DiComm}\\ \textbf{Mode}} & \shortstack{\textbf{Strategy}\\ \textbf{Search}} & \shortstack{\textbf{PP}\\ \textbf{Method}} & \textbf{Optimization}  & \shortstack{\textbf{Relative}\\ \textbf{Iteration Time}} \\
        \hline
        DDR & HeteroAuto & HeteroPP 1F1B & Full & 100\% \\
        \hline
        TCP & HeteroAuto & HeteroPP 1F1B  & Full & 110.1\% \\
        \hline
        DDR & HeteroAuto & Uniform 1F1B & Full & 126.4\% \\
        \hline
        DDR & HeteroAuto & HeteroPP 1F1B  & \shortstack{w/o SR\&AG resharding} & 104.8\% \\
        \hline
        DDR & HeteroAuto & HeteroPP 1F1B  & \shortstack{w/o fine-grained overlap} & 101.8\% \\
        \hline
    \end{tabular}%
    }
    \label{tab:ablation}
\end{table}

\section{Conclusion}
In this paper, we propose a unified framework H2 for efficient large language model training on hyper-heter-ogeneous clusters comprising over 1,000 chips with diverse architectures and software stacks. By integrating DiTorch and DiComm to unify the programming interface and communication library, our approach overcomes the challenges related to operator and communication isolation. In addition, the HeteroPP framework, together with the HeteroAuto automatic strategy search, dynamically optimizes parallelism configurations based on rigorous cost modeling and hardware profiling. Moreover, our topology-aware activation resharding strategy and fine-grained computation-communication overlapping minimizes cross-node data transfer delays, further improving overall performance. Experimental results on a 100B-parameter model demonstrate that our solution achieves a superlinear speedup compared to traditional homogeneous methods by effectively balancing computational loads, memory disparities, and communication overhead.


\bibliographystyle{ACM-Reference-Format}
\bibliography{main}

\appendix

\end{document}